\documentclass[12pt, letterpaper]{JHEP3}

\usepackage{amssymb, amsmath, amsopn, amsthm}
\usepackage{epsfig}

\title{Lectures on Nongeometric Flux Compactifications}

\author{Brian Wecht \\
{Center for Theoretical Physics} \\ {Massachusetts Institute of Technology} \\
{Cambridge, MA 02139, U.S.A.} \\ {bwecht\ {\rm at}\ mit.edu}
}

\abstract{These notes present a pedagogical review of nongeometric flux compactifications. We begin by reviewing well-known geometric flux compactifications in Type II string theory, and argue that one must include nongeometric ``fluxes" in order to have a superpotential which is invariant under T-duality. Additionally, we discuss some elementary aspects of the worldsheet description of nongeometric backgrounds.  This review is based on lectures given at the 2007 RTN Winter School at CERN. }

\preprint{MIT-CTP-3850}

\newcommand{\half}{\frac{1}{2}}

\newcommand{\cD}{\mathcal{D}}

\newcommand{\F}{\mathcal{F}}

\newcommand{\cH}{\mathcal{H}}

\newcommand{\cJ}{\mathcal{J}}

\newcommand{\cL}{\mathcal{L}}

\newcommand{\cN}{\mathcal{N}}

\newcommand{\bZ}{\mathbb{Z}}
\newcommand{\bR}{\mathbb{R}}

\newcommand{\bX}{\mathbb{X}}

\newcommand{\contract}{\llcorner}

\renewcommand{\Re}{{\rm Re}\,}
\renewcommand{\Im}{{\rm Im}\,}

\newcommand{\beq}{\begin{equation}}
\newcommand{\eeq}{\end{equation}}

\newcommand{\er}[1]{(\ref{eq:#1})}

\begin{document}
\section{Introduction}
\label{sec:}

The goal of these lectures is simply to present nongeometric flux compactifications as a natural and essential ingredient in the set of all string compactifications. In some sense, this statement is completely obvious: From the point of view of a two-dimensional conformal field theory, there is no compelling reason that there should be a target space with any conventional geometric interpretation; one expects that relatively few string compactifications will use something as mundane as a manifold. It is not surprising, however, that the string compactifications that have heretofore been the most extensively studied have been those with geometrical target spaces, such as Calabi-Yau manifolds. The study of geometric compactifications is tractable due to the large number of powerful mathematical tools that have been developed, and the study of Calabi-Yau compactifications has yielded many amazing new insights into both string theory and mathematics. 

In the above sense,  the term ``nongeometric" is almost completely devoid of content -- one may as well say ``all string compactifications, except a set of (probably) measure zero." However, here we will not use the term in this very general setting. Instead, we will consider a very specific set of nongeometric compactifications. In particular, we will focus on those that use structures which show up via doing T-duality on a background with NS-NS 3-form flux. Although this is again probably just a very small part of the landscape of string theory vacua, it is a good place to start: We can take a conventional geometric compactification, and ask how T-duality (or mirror symmetry) acts on it. Other kinds of well-studied nongeometric compactifications include Landau-Ginzburg models (see e.g. \cite{Becker:2006ks} for some recent work), and also asymmetric orbifolds, which we will briefly discuss in these notes.

Although one can certainly just focus on NS-NS fluxes and their duals, it is more interesting (and often necessary) to consider backgrounds with both NS-NS and R-R fluxes. In the purely geometric case, such flux compactifications have been of much recent interest in the literature, and were pioneered in \cite{flux-c,gvw,drs}. 
In addition to being interesting in their own right, these backgrounds have the added advantage of stabilizing many (and sometimes all) moduli \cite{gkp,kst,Frey:2002hf}. Although most work on such compactifications has been restricted to Calabi-Yau manifolds, there is an ever-increasing amount of literature focusing on more general geometries, such as G-structure compactifications in both the Type II  \cite{Grana:2006hr,gstrii} and heterotic \cite{gstrhet} string, as well as more generally \cite{gstr}. Additionally, there has been much recent work on generalized geometries  \cite{gengeo}, which were pioneered in the work of Hitchin \cite{hitchin}. 

The fact that performing T-duality on NS-NS fluxes yields nongeometric backgrounds is not a new observation. It was argued in \cite{Kachru:2002sk} that nongeometric spaces show up as duals of known flux compactifications. As we will see in these notes, perfoming one T-duality on a background with H-flux yields a twisted torus compactification, which is characterized by a set of ``geometric fluxes" $f^a_{bc}$. Twisted tori have been studied for many years now \cite{Scherk:1979zr,early-ss-reduction,km}, especially in the context of flux compactifications \cite{ttflux,hre,acfi,vz1,cfi,Grana:2006kf,Villadoro:2007yq,dkpz}; one can similarly twist other types of geometries, as in \cite{cts}. Performing another T-duality on a twisted torus yields a locally geometric but globally nongeometric space, which we will characterize by a nongeometric flux $Q^{ab}_c$. These spaces are globally nongeometric because they are patched together using general transition functions in the perturbative T-duality group. This type of space has been studied both within the context of flux compactifications \cite{ng,stw1,stw2}, and on its own \cite{ao2,x2,othertfold,supertfold,quantumtfold,Dabholkar:2005ve}.

Before we begin, let us briefly describe some caveats. Everything in these notes is done to lowest order in $g_s$ and $\alpha^\prime$; including corrections is no doubt important, and could drastically alter some of the conclusions below. The structure of this review largely follows the presentation in \cite{stw1,stw2}, while making an attempt to be more pedagogical when describing some of the background material. However, the primary purpose of these notes is to emphasize nongeometric backgrounds, and  we will make no attempts to be exhaustive when it comes to vast extant literature on flux compactifications. Finally, we will strive here to emphasize important points rather than arduous calculations, and at times some arguments will be schematic. When we are less than rigorous, we will call attention to it, so that the reader is not misled. 

These notes are organized as follows. In Section 2, we describe Type IIB string theory compactified on a $T^6/\bZ_2$ orientifold. This serves to define some basic notation, as well as provide an example we will use to motivate more complicated compactifications later on. Section 3 discusses Type IIA string theory on a twisted torus, and includes an elementary example of a three-dimensional twisted torus. In Section 4, we review how T-duality acts on string theory backgrounds, and use this to define a series of geometric and nongeometric fluxes. In Section 5, we motivate a superpotential (first derived in \cite{stw1}) which incorporates the compactifications in Sections 2 and 3, and also includes nongeometric fluxes. This superpotential is manifestly invariant under T-duality, and we discuss some properties of the distribution of these vacua, as well as some caveats. In Section 6, we very briefly discuss two different approaches for describing nongeometric spaces via the worldsheet: Asymmetric orbifolds, and Hull's \cite{x2} doubled torus formalism. Finally, in Section 7 we conclude with a discussion of possible directions for future work.

\section{Type IIB on a $T^6/\bZ_2$ orientifold}

Flux compactifications have been a topic of intense study for many years now \cite{flux-c, gvw, drs}, especially since the pioneering work of \cite{gkp}. These lectures do not attempt to be a comprehensive review of flux compactifications, so we will content ourselves here with reviewing material as it is needed. For some more detailed accounts of the general study of flux compactifications, see the reviews in \cite{fluxrev}.

To begin, let us review some well-known facts about Type IIB flux compactifications. The Type IIB Einstein frame supergravity action, which we write mostly just to establish some notation, is
\begin{equation}
S_{IIB} = \frac{1}{2\kappa_{10}^2} \int d^{10}x \sqrt{-g} \left (R - \frac{\partial_M S \partial^M \bar S}{2(\Im S)^2} 
- \frac{G_3 \cdot \bar G_3}{2 \cdot 3! \, \Im S} - \frac{\widetilde F_5^2}{4 \cdot 5!} \right ) + \frac{1}{2\kappa_{10}^2} \int
\frac{C_4 \wedge G_3 \wedge \bar G_3}{4 i \, \Im S},
\label{iibaction}
\end{equation}
where the R-R 0-form and string coupling are combined into the axiodilation $S = C_0 + i e^{-\phi} = C_0 + \frac{i}{g_s}$, and the R-R and NS-NS 3-forms are combined into $G_3 = F_3 - S H_3$. Additionally, $F_3 = dC_2$, $H_3 = dB_2$, and $\widetilde F_5 = dC_4 - \half C_2 \wedge H_3 + \half F_3 \wedge B_2$. As usual, one must impose $ * \widetilde F_5 = \widetilde F_5$ on any solutions, and $\widetilde F_5$ has the standard nonstandard Bianchi identity $d\widetilde F_5 = H_3 \wedge F_3$.

As the authors of \cite{gkp} pointed out, if one wants to turn on nonzero 3-form fluxes wrapping cycles in some six-dimensional compact space, it is necessary to include negative tension objects in order to evade a well-known no-go theorem \cite{de Wit:1986xg,Maldacena:2000mw}. It is for this reason that we will from now on focus exclusively on orientifolds, and in particular (in these lectures) we will only consider IIB compactifications which include O3-planes transverse to the compact directions. 

Let's now work with one of the simplest possible flux compactifications: a $T^6/\bZ_2$ orientifold. This theory has been extensively studied, starting with the work of \cite{kst}, and continuing in e.g. \cite{Kachru:2002sk,iib}. Let the coordinates on the $T^6$ be $x^i$, $i=1,...,6$, such that each coordinate has period 1, $x^i \sim x^i + 1$. We orientifold this $T^6$ by the action of $\Omega \bZ_2 (-1)^{F_L}$, where $\Omega$ is worldsheet parity, $\bZ_2$ takes $x^i  \rightarrow -x^i$, and $(-1)^{F_L}$ gives a $-1$ to any field that comes from a left-moving R sector (i.e. any R-R or R-NS fields). Although the $(-1)^{F_L}$ factor may seem strange, one can show that without it, one does not preserve any SUSY. 

This orientifold breaks half of the 32 supersymmetries, leaving $\cN = 4$ in four dimensions. The bosonic fields in the four dimensional theory are a graviton, 12 vector gauge bosons, and 38 real scalars.  These group into one $\cN=4$ gravity multiplet (a graviton, six vectors, and one complex scalar), and six $\cN=4$ massless vector multiplets (each with a vector and three complex scalars). Since we will want to turn on flux that breaks $\cN = 4$ to $\cN =1$, we should write the $\cN=4$ multiplets in $\cN=1$ language; see Table 1.
\begin{center}
\begin{tabular}{|c|c|}
\hline
$\cN=4$ & $\cN=1$ \cr
\hline
1 gravity  & 1 gravity, 3 spin 3/2, 3 vector, 1 chiral \cr
1 vector & 1 vector, 3 chiral \cr
\hline
\end{tabular}
\vskip0.5cm
{\small {\bf Table 1}:\,\,$\cN=4$ multiplets in $\cN=1$ components.} 
\end{center}

Each $\cN=1$ multiplet has fields with spin $s$ and $s-\half$, as usual. To be explicit, the gravity multiplet has fields with spin 2 and 3/2, the ``spin 3/2" multiplet has fields with spin 3/2 and 1, the vector multiplet has fields with spin 1 and 1/2, and the chiral multiplet has fields with spin 1/2 and 0. As described in \cite{kst}, turning on flux to break $\cN=4$ to $\cN=1$ charges some of the scalars, which then get eaten by gauge bosons; this just comes from the $\widetilde F_5^2$ term in the action. The net result is that generically, one loses  from the massless spectrum twelve of the scalars from R-R 4-form flux. Six of these scalars pair up with each other, and the other six pair up with K\"{a}hler moduli. This leaves twenty scalars, which are in ten chiral multiplets. These ten complex scalars consist of the axiodilation $S$, three K\"{a}hler moduli pairing $C_4$ with the volume of a $T^2$, and six complex structure moduli. 
For our purposes, it is useful to further simplify this to the case where our $T^6 = T^2 \times T^2 \times T^2$, and all the 2-tori are identical. In this case, there are three moduli left in the four-dimensional theory: The axiodilaton $S$, a complex structure modulus $\tau$, and a K\"{a}hler modulus $U$. 

To complete the effective $\cN=1$ supergravity theory, we need to find the superpotential and K\"{a}hler potential. The K\"{a}hler potential is easily derived via dimensional reduction \cite{gkp}: As usual for a $T^2$, the kinetic terms have the $SL(2,\bZ)$ invariant form
\begin{equation}
3\frac{\partial_\mu \tau \partial^\mu \bar \tau}{(\Im \tau)^2} + 3\frac{\partial_\mu U \partial^\mu \bar U}{(\Im U)^2},
\end{equation}
where the factors of 3 come from the three $T^2$'s. The kinetic term for the axiodilaton is unchanged from (\ref{iibaction}). Thus, the combined tree-level K\"{a}hler potential is given by
\begin{equation}
\label{kpot}
K = -3 \ln(-i (\tau -\bar \tau) )- 3 \ln (-i (U - \bar U)) - \ln (-i (S -\bar S)).
\end{equation}

The superpotential is given by the famous Gukov-Vafa-Witten \cite{gvw} formula
\begin{equation}
W = \int G_3 \wedge \Omega = \int (F_3 - S H_3) \wedge \Omega.
\label{gvw} 
\end{equation}
This shows up in the $\cN=1$ supergravity scalar potential \cite{Wess:1992cp}
\begin{equation}
V = e^K \left(\sum_{i,j, = \{\tau, U, S \}} K^{i \bar j} D_i W \overline{D_j
  W} - 3 |W|^2 \right)\,,
\label{eq:scalarpotential}
\end{equation}
where $D_i W = \partial_i W + W \partial_i K$ and $K^{i \bar j}$ is the inverse of the metric $K_{i \bar j} = \partial_i \partial_{\bar j} K$. 
The equations for a supersymmetric vacuum are $D_i W =0$; it is straightforward to check that this implies $\partial_i V = 0$ as well as $\delta \lambda = 0$, where $\lambda$ is a superpartner of any of the moduli. 

Since our compact space is so simple, we can straightforwardly rewrite (\ref{gvw}). Let us define each $T^2$ by a pair of coordinates, with one Greek and one Latin index: $(\alpha, i), (\beta, j), (\gamma, k)$. Throughout the rest of these notes, we will use this notation, where a given Latin or Greek letter stands for a specific coordinate. We can define a complex coordinate on each $T^2$, e.g. $z^1 = x^\alpha + \tau x^i$. This then allows us to write the holomorphic 3-form $\Omega$ as
\begin{eqnarray}
\Omega &=& dz^1 \wedge dz^2 \wedge dz^3 
=  dx^\alpha \wedge dx^\beta \wedge dx^\gamma + 
\tau (dx^\alpha \wedge dx^\beta \wedge dx^k + \cdots  )  \cr
&\qquad& + \tau^2 (dx^\alpha \wedge dx^j \wedge dx^k + \cdots) + 
\tau^3 dx^i \wedge dx^j \wedge dx^k.
\end{eqnarray}
One can now easily do the integral (\ref{gvw}), which gives
\begin{equation}
W = P_1(\tau) + SP_2(\tau),
\end{equation}
where $P_1(\tau)$ is a cubic polynomial in $\tau$ whose coefficients are R-R fluxes $F_3$ integrated over particular cycles, and $P_2(\tau)$ is a cubic polynomial whose coefficients are integrated NS-NS fluxes $H_3$. These coefficients are even integers due to a quantization condition \cite{kst}. 
To be precise, we can write the superpotential as 
\begin{equation}
W  =  a_0 - 3a_1 \tau + 3a_2 \tau^2 - a_3 \tau^3+ S (-b_0 + 3b_1 \tau - 3b_2 \tau^2 + b_3 \tau^3),
\label{wiib}
\end{equation}
where the coefficients are given in Table 2.
\begin{center}
\begin{tabular}{ || c || c || c ||}
\hline
\hline
Term &IIB flux & integer flux\\
\hline
\hline
$1 $&  $ \bar{F}_{ ijk} $& $  a_0 $\\
\hline
$\tau $& $\bar{F}_{ ij \gamma} $& $   a_1 $\\
\hline
$\tau^2 $& $\bar{F}_{i \beta \gamma} $& $  a_2 $\\
\hline
$\tau^3 $& $\bar{F}_{\alpha \beta \gamma} $& $  a_3 $\\
\hline
$S $& $ \bar{H}_{ ijk} $& $  b_0$\\
\hline
$S\tau $& $ \bar{H}_{\alpha jk} $& $  b_1 $\\
\hline
$S \tau^2 $& $ \bar{H}_{ i \beta \gamma} $& $  b_2 $\\
\hline
$S\tau^3 $& $ \bar{H}_{\alpha \beta \gamma} $& $  b_3 $\\
\hline
\hline
\end{tabular}
\vskip0.5cm
{\small {\bf Table 2}:\,\,Fluxes in the IIB superpotential.} 
\end{center}
We have used the notation $\bar F_{abc}$ to indicate the component of $F_3$ integrated over the $abc$-cycle. Since we have taken all the $T^2$'s to be the same, we have only written one representative flux for any given combination of Latin and Greek indices. For example, $\bar F_{ij \gamma} = \bar F_{jk \alpha} = \bar F_{ki \beta}$, and we have only written the first of these in the table. We will use this notation (and assumption) throughout the rest of these notes. 

It is worth noting that the superpotential does not depend on the K\"{a}hler modulus $U$; this is the famous ``no-scale" structure of IIB compactifications. As a result, it appears that (to the approximation we are working here, ignoring e.g. non-perturbative corrections as in \cite{Kachru:2003aw}), we can only fix two of the three moduli. Additionally, the no-scale structure gives a cancellation between some terms in the scalar potential, implying that
\begin{equation}
V = e^K \left(\sum_{i,j, = \{\tau, S \}} K^{i \bar j} D_i W \overline{D_j W}  \right)\, \geq 0.
\end{equation}
In these models, we will not be able to obtain supersymmetric AdS vacua, only Minkowski. 

We have not quite completed our description of the theory, since there is a constraint our fluxes must satisfy. The Chern-Simons type term in the action gives a tadpole for $C_4$, which will get additional contributions from the O3-planes. Said another way, the integrated Bianchi identity for $\widetilde F_5$ implies that
\begin{equation} 
\frac{1}{2\kappa_{10}^2 T_3}\int H_3 \wedge F_3  = 16,
\end{equation}
where $T_3$ is the tension of a D3-brane. The 16 in this formula comes from the $2^6 = 64$ O3-planes we have, each with charge -1/4 of a D3-brane. In terms of our integer fluxes $a_i$ and $b_i$ in (\ref{wiib}), this condition translates directly to 
\beq
a_0 b_3-3 a_1 b_2+3 a_2b_1-a_3b_0 = 16.
\eeq
It is clear what one must do to now find supersymmetric vacua: Pick a set of fluxes consistent with this constraint, and then solve $D_S W = D_\tau W =0$. For specific examples, see \cite{kst,Kachru:2002sk,iib}.

\section{Type IIA on a twisted torus}

In this section, we'll consider IIA compactified on a slightly more complicated space, a twisted torus. This theory has been studied in \cite{Kachru:2002sk, hre, acfi, vz1, cfi,Grana:2006kf,Villadoro:2007yq,dkpz}. First, however, let's start with IIA on an already familiar space, the diagonal and symmetric $T^6$ we considered in the previous section. To see the relationship between this compactification and its IIB counterpart, let's begin by T-dualizing the $T^2 \times T^2 \times T^2$ three times, once one leg of each $T^2$, i.e. in all the Greek directions. We will denote this chain of three T-dualities as $T_{\alpha \beta \gamma}$. This turns the IIB O3-planes into IIA O6-planes which fill the noncompact directions as well as the three Greek directions. 

As in the IIB theory, there are three moduli $S,\tau, U$. These are given in Table 3.\footnote{This table is schematic, and the situation is slightly more complex. See \cite{dkpz} for a more careful treatment of the relationship of these moduli to ten-dimensional fields.} 
\begin{center} 
\begin{tabular}{|c|c|c|}
\hline
&IIA & IIB \\
\hline
$S$& $C^{(3)}_{\alpha \beta \gamma} + ie^{- \phi}$ & $C_0 + ie^{- \phi}$ \\
\hline
$\tau$ &$B_{\alpha i} + i V$ & Complex structure \\
\hline
$U$ & $C^{(3)}_{ij \gamma} + i \tau_2$ & $C^{(4)}_{\alpha i \beta j} + i V$ \\
\hline
\end{tabular}
\vskip0.5cm
{\small {\bf Table 3}:\,\,$S,\tau,U$ in both IIA and IIB.} 
\end{center}
On the IIA side, our orientifold projection forces all the tori to be rectangular, so the complex structure modulus $U$ pairs up the R-R 3-form with the imaginary part of the complex structure. The metric data from the IIB complex structure modulus becomes a component of the $B$-field, which pairs with the volume of any of the $T^2$; this is the familiar K\"{a}hler modulus for a $T^2$ with $B$-field. 

Interestingly, we can place an additional structure on our $T^6$ which is compatible with the orientifold projection. This additional structure will curve the space in a particular way, turning it into a ``twisted torus." To motivate this idea, let's take a brief detour via a simple example. 

\subsection{A simple twisted torus}

Consider a three-dimensional space with coordinates $(x,y,z)$, and metric
\beq
ds^2 = (dx - f^x_{yz}\,  z  \, dy)^2 + dy^2 + dz^2.
\label{ttmetric}
\eeq
If we wish to compactify this space, we can identify $x \sim x+1$ and $y \sim y+1$ without causing any trouble with (\ref{ttmetric}), but if we additionally want to identify $z \sim z+1$, the metric becomes globally ill-defined. We can compensate for this by also shifting $x$ by $f^x_{yz} \,  dy$. Thus, we compactify this space via the identifications
\beq
(x,y,z) \sim (x+1, y, z) \sim (x,y+1,z) \sim (x + f^x_{yz} y, y, z+ 1).
\eeq
These keep the metric globally well-defined, and produce a space topologically distinct from a $T^3$: a twisted torus. One can easily picture this space as a $T^2$ in the $(x,y)$ directions fibered over an $S^1$ in the $z$ direction. As one goes around the $S^1$ base, the fiber $T^2$ undergoes a shift in complex structure $\tau \rightarrow \tau + f^x_{yz}$. If we want to end up with an equivalent fiber after traversing the $S^1$, we need to ensure that this is an $SL(2,\bZ)$ transformation, so we require $f^x_{yz} \in \bZ$. 

There is a very useful way of thinking about the number $f^x_{yz}$. Define the globally invariant one-forms
\begin{eqnarray}
\eta^x & = & dx - f^x_{yz} \,  z \, dy \\
\eta^ y & = & dy \\ 
\eta^z & = & dz. 
\end{eqnarray}
Clearly, $ d \eta^ y = d \eta^z = 0$, but 
\beq
d \eta^x  = f^x_{yz} dy \wedge dz= f^x_{yz} \eta^y \wedge \eta^z.
\eeq
The $f^x_{yz}$ are just components of the spin connection, by Cartan's structure equations. Additionally, they are the structure constants of a Lie group, which show up as above when the Lie group is viewed as a manifold. In the literature, these $f^x_{yz}$ are often referred to as (geo)metric fluxes, for reasons that will shortly become clear. 

The utility of this vielbein approach is that we can easily generalize it. Consider a manifold with a basis of globally defined one-forms $e^a$. The generalization of the above construction is that we can write
\beq
\label{fdef}
de^a = f^a_{bc} \, e^b \wedge e^c,
\eeq
with the $f^a_{bc}$ are all constant. Note that the requirement that $f^a_{bc}$ be constant is a nontrivial constraint on the manifold. Additionally, the $f^a_{bc}$ must also obey a constraint:
\beq
d^2 e^a = 2 f^a_{b [ c} f^b_{de]} e^d e^e e^c = 0.
\eeq
Therefore,  the $f^a_{bc}$ obey a Jacobi identity $f^a_{b [ c} f^b_{de]} =0$, as the structure constants of a Lie algebra should. 

The study of such manifolds (with constant $f^a_{bc}$) is long and varied, and we will not cover it here. Let us make a few brief comments, though. In principle, we can have constant $f^a_{bc}$ on non-compact manifolds as well, and it is worth asking what the criteria for compactness are. This is a long story, and readers are encourage to check out the excellent review in \cite{Grana:2006kf} as well as the mathematics literature for more details. One thing we can say here, though, is that one criterion is that $f^a_{ab} = 0$ (no sum) is a necessary (but not sufficient) condition for compactness; this comes from requiring $d( \alpha e^1 \wedge ... \wedge e^{d-1}) = 0$ for some constant $\alpha$, without which the volume form would be exact. $f^a_{bc}$ that form a nilpotent algebra automatically satisfy this condition. For compactness, they should additionally be rational (as in our three-dimensional example). Such manifolds are called ``nilmanifolds," or ``twisted tori," since they can generally be shown to be tori fibered over tori. 

\subsection{Back to IIA}

We can consistently add some of these $f^a_{bc}$ to our IIA orientifold. This should be thought of as adding some rigid structure to the background metric, which changes the manifold to be something other than a $T^2 \times T^2 \times T^2$. It's easy to show that the only $f^a_{bc}$ that survive the orientifold projection have an odd number of Greek indices, i.e. $f^\alpha_{jk}, f^i_{\beta k}, f^i_{j \gamma}, f^\alpha_{\beta \gamma}$.
Knowing the metric, one can now explicitly do the dimensional reduction, as in  \cite{hre, acfi, vz1, cfi,Grana:2006kf,Villadoro:2007yq, dkpz}. Here, we will simply state the answer. The K\"{a}hler potential is the same as in (\ref{kpot}), and the superpotential is
\beq
\label{wiia}
W = P_1(\tau) + S P_2(\tau) + UP_3 (\tau),
\eeq
where $P_1$ is cubic and $P_2$ and $P_3$ are linear. Specifically, 
\begin{eqnarray}
W & = & a_0 - 3a_1 \tau + 3a_2 \tau^2 - a_3 \tau^3 + S (-b_0 + 3b_1 \tau) + 3 U (c_0 + (\hat{c}_1 +\check{ c}_1 + \tilde{c}_1) \tau). 
\end{eqnarray}
The coefficients are given in Table 4.
\begin{center}
\begin{tabular}{  || c || c || c ||}
\hline
\hline
Term & IIA flux  & integer flux\\
\hline
\hline
$1 $& $ \bar{F}_{\alpha i \beta j \gamma k}$& $  a_0 $\\
\hline
$\tau $& $ \bar{F}_{\alpha i \beta j} $& $   a_1 $\\
\hline
$\tau^2 $& $ \bar{F}_{\alpha i} $& $  a_2 $\\
\hline
$\tau^3 $& $ F^{(0)}  $& $  a_3 $\\
\hline
$S $& $ \bar{H}_{ijk} $& $  b_0$\\
\hline
$S\tau $& $ f^\alpha_{j k} $& $  b_1 $\\
\hline
$U $& $ \bar{H}_{\alpha \beta k} $& $  c_0 $\\
\hline
$U\tau $& $ f ^ j_{k\alpha}, f^i_{\beta k}, f^\alpha_{\beta \gamma} $&
$\check{c}_1,  \hat{c}_1,\tilde {c}_1 $\\
\hline
\hline
\end{tabular}
\vskip0.5cm
{\small {\bf Table 4}:\,\,Fluxes in the IIA superpotential.} 
\end{center}

Although this construction stands on its own, it will be useful for us to see how some of these terms relate to analogous terms in the IIB superpotential. The coefficients in $P_1(\tau)$ are all the even R-R fluxes, which are just the $T_{\alpha \beta \gamma}$-duals of the R-R fluxes in IIB. The constant term in $P_2(\tau)$ is the same as it was in the IIB superpotential. However, there are some terms here that did not show up in the IIB superpotential, notably everything in $P_3(\tau)$. As a result, note that {\bf all} the moduli appear in this IIA superpotential, and we do not need e.g. nonperturbative effects to stabilize everything. The mismatch between the IIA and IIB superpotentials is something we will return to shortly, and will be the key observation motivating our nongeometric compactifications. 

As in the IIB theory, there are constraints we must satisfy. One constraint comes from $dH=0$, which was automatically satisfied in the IIB model we constructed. Here, however, our one-forms are not closed, so $dH=0$ becomes nontrivial and yields the constraint $\bar H_{a[bc} f^a_{de]} =0$. Additionally, as described previously, the $f^a_{bc}$ must satisfy a Jacobi identity $f^a_{b[c}f^b_{de]} = 0$. 

There are constraints from the R-R sector as well, in the form of integrated Bianchi identities. These are of the form $(d + H)\wedge F = 0$, where $F$ is any of the $F_0, F_2, F_4, F_6$ R-R fluxes. This was the case in IIB as well, where our constraint was from the Bianchi identity for $\widetilde F_5$. In this IIA model, the nontrivial constraints come from $dF_2$ and $dF_4$, which give
\begin{eqnarray}
\bar F_{a [b}f^a_{cd]} + F_0\bar H_{bcd} &=& 0 \\
\bar F_{x[abc}f^x_{de]} + \bar F_{ab} \bar H_{cde]} &=& 0. 
\end{eqnarray}
The first term in each equation is a result of the basis one-forms not being closed. Now, as before, to find supersymmetric solutions, one simply imposes some set of fluxes consistent with these four (two NS-NS and two R-R constraints), and then solves $D_i W = 0$. For some examples of how to solve the equations of motion for IIA vacua, see \cite{iia}.

\section{Fluxes and T-duality}

Let's pause now to consider both of the compactifications just discussed. IIB on a $T^2 \times T^2 \times T^2$ with NS-NS and R-R 3-form fluxes gave a superpotential
\beq
W = P_1(\tau) + SP_2(\tau),
\eeq
with $P_{1,2}$ cubic. IIA on a twisted $T^2 \times T^2 \times T^2$, with 0-,2-,4- and 6-form R-R fluxes and NS-NS 3-form flux gave a superpotential
\beq
W = P_1(\tau) + S P_2(\tau) + UP_3 (\tau),
\eeq
with $P_1$ cubic and $P_{2,3}$ linear. 

At least part of these superpotentials behave nicely under three T-dualities: $P_1$ changes exactly as one would expect, with the R-R 3-form flux in IIB turning into the even-form R-R fluxes in IIA via the T-duality rule \cite{trr}
\beq
\bar F_{x\alpha_1 \cdots \alpha_p}
\stackrel{T_x}{\longleftrightarrow}
\bar F_{\alpha_1 \cdots \alpha_p} \,.
\label{eq:ft}
\eeq
Note that although there are other terms in this T-duality rule, we focus on the (integrated) topological fluxes, which are independent of gauge choices. Similarly, the constant part of $P_2$, which is H-flux wrapped on the non-dualized cycles, does not change, as one would expect. As we will discuss shortly, the linear term in $P_2$ also behaves as expected under duality. However, there is otherwise a mismatch between these two superpotentials. Most notably, the IIB superpotential has no $U$ dependence, and nothing analagous to $P_3$. In order to understand this mismatch, we need to better understand how R-R and NS-NS fluxes behave under T-duality, so let's take a minute to study this further. 

T-duality is a symmetry of string theory which relates string theory on a circle of radius $R$ to string theory on a circle of radius $\alpha^\prime / R$. It is natural to ask how T-duality acts on a more general background, and we can see how as follows: Consider the free bosonic string on a circle, which we'll call the $\theta$ direction. The worldsheet action is
\beq
S = \frac{1}{2\pi} \int d^2 z \,  \partial \theta \, \bar \partial \theta \,  G_{\theta \theta}.
\label{orig}
\eeq
One can introduce a Lagrange multiplier $\widetilde \theta$, writing
\beq
S = \frac{1}{2\pi} \int d^2 z \, \left [ G_{\theta \theta} L \bar L+ \tilde \theta \left ( \partial \bar L - \bar \partial L \right ) \right ].
\eeq
Integrating out $\widetilde \theta$ restores the original action (\ref{orig}), but integrating out $L$ yields
\beq
S = \frac{1}{2\pi} \int d^2 z \,  \partial \widetilde \theta \, \bar \partial \widetilde \theta \,  \frac{1}{G_{\theta \theta}}.
\eeq
Thus we see that the two conformal field theories, one with target space metric $G_{\theta \theta}$ and the other with target space metric  $1/G_{\theta \theta}$,  are classically equivalent. One can also check that these theories are equivalent quantum mechanically \cite{Buscher:1987sk, Buscher:1987qj,Rocek:1991ps}. Notice that it was essential that we had an isometry in the $\theta$ direction, so that the metric was independent of this coordinate.

For a more general background including metric, B-field, and dilaton, one can repeat the above procedure. This was first done by Buscher in \cite{Buscher:1987sk}. The ``Buscher rules" give a new background $G^\prime_{\mu \nu}, B^\prime_{\mu \nu}, \phi^\prime$ in terms of the originial background $G_{\mu \nu}, B_{\mu \nu}, \phi$, after T-dualizing in (say) the $x$ direction:
\begin{eqnarray}
G^\prime_{xx} &=& \frac{1}{G_{xx}}, \qquad
G^\prime_{x \mu}  = -\frac{B_{x \mu}}{G_{xx}}, \qquad
B^\prime_{x \mu} = -\frac{G_{x \mu}}{G_{xx}} \\
G^\prime_{\mu \nu}  &=&  G_{\mu \nu} - \frac{G_{x \mu}G_{x \nu} - B_{x \mu} B_{x \nu}}{G_{xx} } \\
B^\prime_{\mu \nu}  &=&  B_{\mu \nu} - \frac{G_{x \mu}B_{x \nu} - B_{x \mu} G_{x \nu}}{G_{xx} } \\
e^{\phi^\prime} &=& \frac{e^{\phi}}{\sqrt{G_{xx}}}
\label{buscherrules}
\end{eqnarray}
The shift in the dilaton is not a consequence of the classical procedure detailed above, and requires a one-loop calculation. Note that once again, we needed an isometry in the $x$ direction to ensure that we could make the metric and $B$-field independent of this coordinate; otherwise, this procedure is not valid \cite{Rocek:1991ps}. 

\subsection{$T^3$ with $H_3$}

Let's apply the Buscher rules to a simple example, that of a $T^3$ with $H_3$ flux. We note immediately that this background does not satisfy the string equations of motion, since it is a flat background with nontrivial $H$-flux. This is not a problem, however, as we only use this as an illustrative example and one could e.g. fiber this $T^3$ over something else to get a good string background. For a slightly different approach that works through this same example, see \cite{Kachru:2002sk}.

To start, take $(x,y,z)$ as the coordinates on the $T^3$, each with period 1. Additionally, put $N$ units of $H$-flux on the torus, such that 
\beq
\int_{T^3} H_3 = N,
\eeq
where we have set a prefactor of $1/(2\pi)^2 \alpha^\prime =1$ for convenience. To ensure that this quantization condition is satisfied, we can now pick a gauge where $B_{xy} = Nz$, with $N \in \bZ$. We have introduced an explicit dependence on the coordinate $z$; we will comment about this further later on. On can view this space not only as a $T^3$, but also as a $T^2$ in the $(x,y)$ directions fibered over an $S^1$ in the $z$ direction, where the K\"{a}hler modulus $\rho = ( \int B) + iV$ of the $T^2$ undergoes $\rho \rightarrow \rho + N$ as $z \rightarrow z+1$. 

Nothing depends on the coordinates $x$ and $y$, so we can feel free to do a T-duality in either of those directions. T-dualizing on the $x$ direction yields the background
\beq
ds^2 = (dx - N z \, dy)^2 + dy^2 + dz^2 ; \qquad B=0.
\eeq
This is exactly the metric (\ref{ttmetric}) with $f^x_{yz} = N$, so we see that this background is a twisted torus. And as before, in order to make this metric globally well-defined, we need to identify $(x,y,z) \sim (x + Ny, y, z+1)$. Thus, we see that a T-duality takes 
\begin{equation}
H_{xyz} \stackrel{T_x}{\longrightarrow} f^x_{yz}.
\eeq
As before, this is a $T^2$ fibered over an $S^1$, where the complex structure $\tau \rightarrow \tau + N$ as $z \rightarrow z+1$. This is as expected: one T-duality has switched the complex structure and K\"{a}hler moduli. 

We still have another T-duality we can do here, since nothing depends on the $y$ direction.\footnote{Strictly speaking, we should be careful here, since the Killing vectore $\partial /\partial y$ is really no longer globally well-defined. There are many reasons for believing that it should be possible to still do this T-duality, but there is currently no rigorous argument in the literature. See \cite{Hull:2006qs} for one argument; another is that one can work with the covering space where the Killing vector is well-defined, and quotient by a translation plus T-duality.} The Buscher rules give 
\beq
ds^2 = \frac{1}{1+N^2 z^2} (dx^2 + dy^2 ) + dz^2 ; \qquad B_{xy} = \frac{Nz}{1+N^2 z^2}.
\label{qflux}
\eeq
It now looks as if $z \rightarrow z+1$ produces a complete mess, and it is difficult to understand this action from (\ref{qflux}). However, it is easy to see what's happening by examining the K\"{a}hler modulus. One can easily check that
\beq
\frac{1}{\rho} = Nz -i,
\eeq
so $z \rightarrow z+1$ just takes $1/\rho \rightarrow 1/\rho + N$. This is an $SL(2,\bZ)$ transformation on $\rho$, so once again we see the fiber $T^2$ shifting its K\"{a}hler modulus as we go around the base circle. 

This is our first example of a nongeometric background, since the transformation $1/\rho \rightarrow 1/\rho + N$ mixes the metric and the B-field. More precisely, this background is locally geometric, since the metric and B-field are defined at every point, but it is not globally a manifold. Upon going around a cycle, the metric and B-field mix by an $SL(2,\bZ) \subset O(2,2;\bZ)$ transformation. As with the previous two backgrounds, this one is characterized by an integer $N$. Writing
\begin{equation}
H_{xyz} \stackrel{T_x}{\longrightarrow} f^x_{yz}
\stackrel{T_y}{\longrightarrow} Q^{xy}_z,
\eeq
we will say that this background is characterized by the nongeometric flux $Q^{xy}_z$. One can show that this object $Q^{xy}_z$ behaves like a one-form, as expected. See Section 3 of \cite{stw1} for details. 

It now appears that we have exhausted every possible T-duality, since there is clearly not an isometry in the $z$ direction. As we will soon see, however, there must be a sense in which this T-duality exists. We will characterize this final, very mysterious, background by the quantity $R^{xyz}$, thus completing the chain
\begin{equation}
H_{xyz} \stackrel{T_x}{\longrightarrow} f^x_{yz}
\stackrel{T_y}{\longrightarrow} Q^{xy}_z 
\stackrel{T_z}{\longrightarrow} R^{xyz}.
\eeq

Assuming for the moment that there are compelling reasons for believing that this final step exists (which we will justify in the next section), we can hypothesize what kind of space has $R^{xyz}$ flux. We claim that the presence of this flux indicates a lack of even a locally geometric description \cite{stw2}. To see this, consider our original $T^3$ with $H_3$ flux, and wrap a D3-brane on the torus. If we were to perform three T-dualities, we would end up with a space with a D0-brane and $R^{xyz}$ flux. Yet this cannot exist, since we cannot wrap a D3-brane on a $T^3$ with $H_3$ flux; the Bianchi identity for the gauge field on the D-brane will then be nonzero, $dF_2 = H_3 \neq 0$. It seems reasonable to guess that since we cannot have D0-branes on a space with $R^{xyz}$ flux, we have lost any description of this space as having spacetime points. 
It is worth pointing out here that these fluxes have been studied from a more rigorous mathematical standpoint than we take here \cite{math}, where it is argued that the presence of $R^{xyz}$ flux corresponds to some kind of nonassociative geometry. 

\section{A duality-invariant superpotential}

We are now in a position to argue for a superpotential which is invariant under T-duality, as was first derived in \cite{stw1}. Let us start with the IIA orientifold on a twisted torus, as described in Section 3 of these notes. We now take this theory through the following steps:

\begin{enumerate}
\item T-dualize on the three Greek directions, arriving at IIB with O3-planes.
\item Now, since IIB has only O3-planes transverse to the compact space, there is a rotational symmetry between the Latin and Greek directions. So let's rotate each $T^2$ by flipping Latin/Greek indices, $\alpha \leftrightarrow i$. This takes e.g. $ x^\alpha + \tau x^i \leftrightarrow x^i + \tau x^\alpha$, so it exchanges $1 \leftrightarrow \tau^3$ and $\tau \leftrightarrow \tau^2$ in the polynomials $P_{1,2,3}$ in the superpotential. 
\item Finally, T-dualize on the Greek directions to get back to IIA. 
\end{enumerate}
One can thus fill in every box in the Table 5.
\begin{center}
\begin{tabular}{ || c  || c || c || c ||}
\hline
\hline
Term & IIA flux & IIB flux & integer flux\\
\hline
\hline
$1 $& $ \bar{F}_{\alpha i \beta j \gamma k}$& $ \bar{F}_{ ijk} $& $  a_0 $\\
\hline
$\tau $& $ \bar{F}_{\alpha i \beta j} $& $\bar{F}_{ ij \gamma} $& $   a_1 $\\
\hline
$\tau^2 $& $ \bar{F}_{\alpha i} $& $\bar{F}_{i \beta \gamma} $& $  a_2 $\\
\hline
$\tau^3 $& $ F^{(0)}  $& $\bar{F}_{\alpha \beta \gamma} $& $  a_3 $\\
\hline
$S $& $ \bar{H}_{ijk} $& $ \bar{H}_{ ijk} $& $  b_0$\\
\hline
$U $& $ \bar{H}_{\alpha \beta k} $& $  Q^{\alpha \beta}_k $& $  c_0 $\\
\hline
$S\tau $& $ f^\alpha_{j k} $& $ \bar{H}_{\alpha jk} $& $  b_1 $\\
\hline
$U\tau $& $ f ^ j_{k\alpha}, f^i_{\beta k}, f^\alpha_{\beta \gamma} $& $ 
Q ^{\alpha j}_k, Q^{i \beta}_k, Q^{\beta \gamma}_\alpha $&
$\check{c}_1,  \hat{c}_1,\tilde {c}_1 $\\
\hline
\hline
$S \tau^2 $& $ Q^{\alpha \beta}_k $& $ \bar{H}_{ i \beta \gamma} $& $  b_2 $\\
\hline
$U \tau^2 $& $  Q ^{\gamma i}_\beta, Q^{i \beta}_\gamma, Q^{ij}_k $& $ Q ^{i\beta}_\gamma, Q^{\gamma i}_\beta,
Q^{ij}_k $& $\check{c}_2,  \hat{c}_2,\tilde{c}_2 $\\
\hline
$S\tau^3 $& $  R^{\alpha \beta \gamma} $& $ \bar{H}_{\alpha \beta \gamma} $& $  b_3 $\\
\hline
$U \tau^3 $& $ R^{ij \gamma} $& $  Q^{ij}_{\gamma} $& $c_3 $\\
\hline
\hline
\end{tabular}
\vskip0.5cm
{\small {\bf Table 5}:\,\,Fluxes in the duality-invariant superpotential.} 
\end{center}

Thus, we arrive at a superpotential
\begin{equation}
W = P_1 (\tau) + SP_2 (\tau) + UP_3 (\tau),
\label{eq:w-complete}
\end{equation}
or, more explicitly, 
\begin{eqnarray}
W & = & a_0 - 3a_1 \tau + 3a_2 \tau^2 - a_3 \tau^3\\ & &
 \hspace{0.2in} + S (-b_0 + 3b_1 \tau - 3b_2 \tau^2 + b_3 \tau^3)
\nonumber\\ & &
 \hspace{0.2in} + 3 U (c_0 + (\hat{c}_1 +\check{ c}_1 + \tilde{c}_1)
 \tau - (\hat{c}_2 +\check{ c}_2 + \tilde{c}_2) \tau^2 -c_3
 \tau^3), \nonumber
\end{eqnarray}
where all the polynomials are cubic in $\tau$, with the coefficients $a_i,b_i,c_i$ defined in Table 5. This is a complete, duality-invariant superpotential. One can additionally relax the condition that all the $T^2$'s be identical, and obtain a superpotential which reduces to (\ref{eq:w-complete}) upon assuming this extra symmetry \cite{acfi}. Note that all we did to motivate this superpotential was take a perfectly good geometric theory (IIA on a twisted torus), and manipulate it via T-duality and rotational symmetry. This produces a superpotential with coefficients that are nongeometric fluxes, including (shockingly) the very mysterious object $R^{xyz}$. 

We can make some additional comments here. Strictly speaking, we have not argued that one can turn on every term in the full superpotential (\ref{eq:w-complete}) at the same time. Rather, our argument is really only for a subset of its terms (corresponding to the original terms turned on in the IIA twisted torus compactification). It is possible that one cannot simply turn on all the terms in this superpotential. Assuming that we can turn on all the terms, however, the general superpotential appears to have no duality frame in which the compactification is geometric. Since all moduli appear in the superpotential, such a theory will generically stabilize all moduli. This can be confirmed numerically, and we will discuss solutions later on in these lectures. Finally, we note that it was shown in \cite{Grana:2006hr} that one can use $SU(3) \times  SU(3)$-structure compatifications to reproduce this superpotential; these compactifications are also generically nongeometric. This independent result is one reason to believe that one might be able to turn on all the terms in (\ref{eq:w-complete}).

The equations of motion are $D_i W = 0$, which are simply 
\begin{eqnarray}
\nonumber
P_1 (\tau) + \bar{S}P_2 (\tau) +UP_3 (\tau)& = &  0\\
\label{eom}
P_1 (\tau) + S P_2 (\tau) + \frac{1}{3}(2U+\bar{U})P_3 (\tau) & = &  0 \\
\nonumber
(\tau -\bar{\tau})  \partial_\tau W -W  &=& 0.
\end{eqnarray}
A detailed analysis of the solutions coming from these equations is a difficult task, but we can make a some general comments without resorting to numerics. For example, consider a geometric IIA compactification, which means that $P_2(\tau)$ is linear, $P_2(\tau) = a + b\tau$ for some $a$ and $b$. If $W=0$ (i.e. a Minkowski vacuum), then the first equation in 
(\ref{eom}) implies $P_2(\tau)=0$. Therefore $\tau \in \bR$, which is a degenerate zero-volume case. Therefore, IIA Minkowski theories (in this model) must be nongeometric. This agrees with a no-go theorem in \cite{Micu:2007rd}, which says it is impossible to get SUSY Minkowski vacua with all moduli stabilized in IIA without nongeometric fluxes. 

\subsection{Constraints}

As before, there are constraints on the fluxes we must satisfy. Rather than belabor the point, we will simply write these down, and provide only a brief explanation. To see the constraints on the (integrated) R-R fluxes, we start with the Bianchi identity in the absence of localized sources $d\widetilde F_5= \bar F_{[abc} \bar H_{def]} = 0$. Since this has six uncontracted indices, it is easy to check that there are no additional terms (using $f, Q,$ or $R$) that could be incorporated here. We can now proceed by T-duality. For example, one T-duality gives the constraint
\beq
\label{1t}
\bar{F}_{x [abc}f^x_{ de]}- \bar{F}_{[ab} \bar{H}_{cde]}  =  0.
\eeq
The first term in (\ref{1t}) arises from components of $F_{abc}$ not in the dualized direction, and the second term comes from the components in the dualized directions. This constraint is exactly $(d + H) \wedge F_4 = 0$, as expected in IIA. 

One can now continue T-dualizing to work out the rest of the R-R constraints, which are
\begin{eqnarray}
\bar{F}_{[abc} \bar{H}_{def]} & = &  0\label{eq:R-R1}\\
\bar{F}_{x [abc}f^x_{ de]}- \bar{F}_{[ab} \bar{H}_{cde]} & = &  0\label{eq:R-R2}\\
\bar{F}_{xy [a b c}Q^{xy}_{d]}
 -3 \bar{F}_{x[ab}f^x_{cd]}- 2\bar{F}_{[a} \bar{H}_{ bcd]} & = &  0\label{eq:R-R3}\\
\bar{F}_{xyz[a b c]} R^{xyz}  -9\bar{F}_{xy [a b}Q^{xy}_{c]}
 -18 \bar{F}_{x[a}f^x_{ bc]}+6F^{(0)}\bar{H}_{[abc]} & = &  0 \label{eq:R-R4}\\
\bar{F}_{xyz[a b]} R^{xyz}  + 6\bar{F}_{xy  [a}Q^{xy}_{b]}
-6\bar{F}_xf^x_{[ab]} & = &
0
\label{eq:R-R5}\\
\bar{F}_{xyza} R^{xyz}  -3\bar{F}_{xy}Q^{xy}_a & = &  0\label{eq:R-R6}\\
\bar{F}_{xyz} R^{xyz} & = &  0 . \label{eq:R-R7}
\end{eqnarray}
These constraints are simply $dF_i =0$ in the presence of geometric and nongeometric fluxes, where $i$ runs from 5 (\ref{eq:R-R1}) to 1 (\ref{eq:R-R6}). The final constraint is a bit mysterious, but is required by duality. In the presence of localized sources, the RHS of these equations would be nonzero. 

There are also constraints on the NS-NS fluxes, which we can obtain by starting with $dH=f ^ x_{[ab} \bar{H}_{cd]x} = 0$ and dualizing. 
T-duality then gives the set of NS-NS constraints
\begin{eqnarray}
\label{eq:ns-1}
\bar{H}_{x[ab}f^x_{cd]} & = &  0\\
\label{eq:ns-2}
f^a_{x[b}f^x_{cd]} + \bar{H}_{x[bc}Q^{ax}_{d]} & = &  0\\
\label{eq:ns-3}
Q^{[ab]}_xf^x_{[cd]}-4f^{[a}_{x[c}Q^{b]x}_{d]} + \bar{H}_{x [cd]} R^{[ab]x} & = &  0\\
\label{eq:ns-4}
Q^{[ab}_xQ^{c]x}_d + f^{[a}_{xd} R^{bc]x} & = &  0\\
\label{eq:ns-5}
Q^{[ab}_x R^{cd]x} & = &  0.
\end{eqnarray}
Additionally, for the $f$- and $Q$-fluxes to be individually T-dual to $H$-flux, they should satisfy 
\beq
f ^ x_{xa} = 0 = Q ^{ax}_x.
\eeq
This constraint on $f^a_{bc}$ is one of the requirements for compactness, as discussed in Section 3.1.

The NS-NS constraints have a very nice interpretation. A  ten-dimensional theory reduced on a $T^6$ gives a four-dimensional theory with a $U(1)^{12}$ gauge group, with 6 generators $Z_m$ ($m=1,...6$) corresponding to diffeomorphisms, and the other 6 generators $X^m$ corresponding to shifts of the B-field. As shown in \cite{Scherk:1979zr,km}, turning on $H$-flux and geometric flux makes the Lie algebra non-abelian, with the fluxes acting as structure constants:
\begin{eqnarray}
{[Z_a, Z_b]} & = & \bar{H}_{abc} X^ c+f ^ c_{ab} Z_c\\
{[Z_a, X ^ b]} & = &-f ^ b_{ac} X ^ c \\ 
{[X ^ a, X ^ b]} & = & 0.
\end{eqnarray}

These equations are not closed under T-duality, but it is easy to fill in the gaps:
\begin{eqnarray}
\label{eq:commute-1}
{[Z_a, Z_b]} & = & \bar{H}_{abc} X^ c+f ^ c_{ab} Z_c\\
\label{eq:commute-2}
{[Z_a, X ^ b] } & = &-f ^ b_{ac} X ^ c +Q ^{bc}_aZ_c\\
\label{eq:commute-3}
{[X ^ a, X ^ b]} & = & Q ^{ab}_c X ^ c +R ^{abc} Z_c.
\end{eqnarray}
The nongeometric fluxes also show up as structure constants in the four-dimensional gauge group! From this perspective, one expects that a general  $\cN=1$ gauged supergravity will lift to a nongeometric compactification. The study of such gauged supergravities, and their string lifts, is a topic of ongoing study \cite{gauged}.

We will also mention here that both the NS-NS and R-R constraints can be packaged into an even more compact form \cite{stw2}. Define a generalized derivative operator
\beq
\cD\equiv H\wedge +f\cdot+Q\cdot+R\contract,
\eeq
where $f,Q$, and $R$ act by contracting all upper indices and antisymmetrizing all 
uncontracted lower indices.
The R-R constraints can now be written as 
\beq
\cD\bar \F = 0,
\eeq  
where $\bar \F$ is a formal sum on all the integrated R-R fluxes.
Additionally, the NS-NS
constraints are equivalent to
\beq
\label{eq:constraints-NS}
\cD ^ 2 = (H\wedge +f\cdot+Q\cdot+R\contract) ^ 2 = 0.
\eeq
In addition to the generalized Bianchi identities, \er{constraints-NS} also implies the
 conditions $f ^ x_{xa} = 0 = Q ^{xa}_x$.

\subsection{Solutions}

In light of (\ref{eq:commute-1}) - (\ref{eq:commute-3}), it appears that the string lift of $\cN =1$ supergravity with a generic rank 12 gauge group is nongeometric, even just in the specialized model we are considering here. It is thus worth investigating the (supersymmetric) solutions to this theory. Although we can simply things a bit analytically, we must solve the equations of motion (\ref{eom}) numerically. This was done in \cite{stw2}, and we will now say a few words about these solutions for supersymmetric vacua. 

The general procedure is, as before, to find sets of fluxes satisfying the constraints, and then solve the equations of motion. There are 14 distinct fluxes, subject to 7 distinct constraints, so even just finding sets of fluxes that solve the constraints is a nontrivial task\footnote{Although Table \cite{stw2} indicates more than 14 fluxes, it turns out that $\hat c_1 = \check c_1$ and $\hat c_2 = \check c_2$, so there are only 14 independent parameters. Similarly, not all the constraints are independent, and some are identically zero in the model we consider here.}. One can analytically solve the equations of motion for $S$ and $U$, leaving only a complicated polynomial in $\tau$ which must generically be solved numerically. 

This procedure was implemented in \cite{stw2}, and we now summarize some broad features of the solutions. It is useful to describe solutions by the string coupling $g_s = 1/\Im S$ and cosmological constant $\Lambda = -3e^K |W|^2$, since these quantities are of physical interest. For vacua to be even potentially trustworthy, both of these quantities should be small: We want to be able to trust our perturbative approximation and ensure that there are not additional light modes we have not included. Having small $g_s$ and $\Lambda$ is clearly not a necessary and sufficient condition for having a trustworthy vacuum, however, and in fact it is not generally clear when the solutions we describe are actually valid. However, we will treat small $(g_s, \Lambda)$ as a rough criterion for good solutions herein.

There is numerical evidence that there are an infinite number of solutions in {\bf any} finite region of $(g_s, \Lambda)$-space, even for regions where the parameters are not small. In  \cite{stw2} the authors arrived at this surprising conclusion by taking sets of fluxes within some bounded region, and solving the equations of motion. As the size of the fluxes was increased, formerly populated regions of $(g_s,\Lambda)$-space became more densely populated, while the overall size of the populated region increased. If this behavior continued to arbitrarily large fluxes, any finite region of $(g_s,\Lambda)$-space would eventually become densely populated. This conclusion is purely numeric, however, and it would be nice to see firmer analysis of the distribution of these vacua, as in \cite{stats}.

 In general, the size of $g_s$ is correlated with the size of $\Lambda$, since $\Lambda \sim e^K \sim e^{-\ln \Im S} \sim g_s$. That we can easily get large string coupling and cosmological constant is somewhat depressing, and there is of course no reason to think that solutions with large values of these parameters are reliable. However, there are many infinite families of solutions which give arbitrarily small $(g_s,\Lambda)$. These solutions were presented in  \cite{stw2}, where the authors showed that there are an infinite number of solutions with tunably small $(g_s, \Lambda)$. One would hope, then, that this might be the place to look for good nongeometric theories. 

One final thing worth pointing out is a very large degeneracy in these solutions. Consider a real shift of the modulus $U$, $U \rightarrow U+1/n$ for some $n \in \bZ$. 
The superpotential $W = P_1 + SP_2 + UP_3$ can then be kept unchanged by taking $P_3 \rightarrow nP_3$ and $P_1 \rightarrow P_1 - P_3$. This is always possible and does not spoil constraints, as one can easily check. Since $W$ and $\Im U$ are unchanged, but $\Re U$ is different, this gives a physically distinct vacuum with the same value of $g_s$ and $\Lambda$. As the coefficients in $P_3$ get larger, we can do more such shifts, making this degeneracy arbitrarily large. Although one might hope that these solutions are physically equivalent, there is no known symmetry that relates them. 

Thus, there appears to be evidence for two types of infinities among the number of nongeometric vacua: One due to a degeneracy in $(g_s, \Lambda)$-space, and the other due to finite regions of  $(g_s, \Lambda)$-space becoming densely populated. If these solutions are to be trusted, then we have found an infinite number of nongeometric vacua, even for fixed tadpole. 

There are many reasons one might doubt this claim, however. Let us make a brief  list of why there might not actually be an infinite number of nongeometric vacua. 
\begin{enumerate}
\item Clearly, arbitrary regions of $(g_s,\Lambda)$-space will contain regions with large string coupling. These solutions are clearly not to be trusted, since the theory is wildly out of perturbative control. 
\item Because our nongeometric spaces are patched together with general T-duality symmetries, one expects that the solutions will be generically string scale. We thus lose any large volume approximation, and have no control over $\alpha^\prime$ effects. 
\item Our infinities come from turning on increasingly large fluxes, and we have not considered backreactions. This is clearly important, and may destroy many of these solutions.
\item Our arguments are entirely from the four-dimensional effective theory. It is possible that we cannot lift all, or even most, of these vacua to full string theory descriptions. This is a pressing problem in the study of nongeometric backgrounds.
\item Even if we could construct string theories from these backgrounds, one would need to check e.g. modular invariance. As we will soon see, some nongeometric backgrounds are related to asymmetric orbifolds, for which modular invariance is notoriously delicate \cite{ao2,ao,Aoki:2004sm,Hellerman:2006tx}.
\item The degeneracy we have found from shifting $U \rightarrow U + 1/n$ is suggestive of a modular redundancy,  but no known modular transformation exists that would relate these solutions. However, it may be possible that a heretofore unknown redundancy removes this degeneracy.
\item Finally, we note that it is unclear that we have consistently truncated the theory to include all the light modes. Other light modes may exist, and as such would need to be included to trust this four-dimensional description. In order to better understand this, it is necessary to lift these solutions to string theory.  
\end{enumerate}
Despite these caveats, it is this author's opinion that it is still very much worth studying these theories, since we can hope to discover some generic features of such vacua while we continue working to rigorously justify our actions via a full string theory description.

\section{Worldsheet Descriptions}

The study of how to rigorously define nongeometric string theories is something which is still an active area of research. The purpose of this section is just to provide some basic ideas, which the interested reader can then go explore further in the references. We first note that the techniques described below may not describe the backgrounds discussed previously in these notes; we present them here purely as examples of how to describe nongeometric target spaces on the worldsheet.

One avenue of study is the relationship of nongeometric theories to asymmetric orbifolds. In general, asymmetric orbifolds are quite difficult to work with, and it is no easy task to find a putative orbifold for a given nongeometric background. Here, we will content ourselves with describing a very simple such orbifold, and show that it describes a nongeometric background. 

The basic idea behind an asymmetric orbifold is simple \cite{ao}: As usual, break the target space coordinate $X^\mu (z, \bar z)$ into left-moving and right-moving parts, writing $X^\mu(z, \bar z) = X_L(z) + X_R (\bar z)$.  An asymmetric orbifold is an orbifold which treats the left- and right-moving parts separately, orbifolding each by a different group. Because of this strange action on the spacetime coordinate, it is reasonable to guess that a generic asymmetric orbifold will be nongeometric. This connection between nongeometric backgrounds and asymmetric orbifolds has been studied in \cite{ao2,Hellerman:2006tx}. 

Let's build up a simple example by starting with a standard orbifold. Consider, in the bosonic string, a $T^2$ fibered over an $S^1$, where as usual we will take the coordinate on the $S^1$ to be $z$. We can orbifold by a translation plus a shift, given by the action
\beq
\tau \rightarrow -\frac{1}{\tau} \quad {\rm as} \quad z \rightarrow z+1,
\eeq
where $\tau$ is the complex structure of the $T^2$ fiber. This is a fine orbifold, and is modular invariant. In fact, $\tau \rightarrow -1/\tau$ is just a $90^\circ$ rotation of the fiber torus. The shift is there in order to make the twisted sectors massive, as well as to induce a Sherk-Schwarz-type potential for the moduli. 

We can T-dualize along one leg of the fiber $T^2$, which switches $\tau \leftrightarrow \rho$. Thus, we have a theory orbifolded by the action
\beq
\rho \rightarrow -\frac{1}{\rho} \quad {\rm as} \quad z \rightarrow z+1.
\eeq
This is a nongeometric background, since the action $\rho \rightarrow -1/\rho$ mixes the metric and $B$-field (recall that $\rho = B + iV$). It's easy to see that this is an asymmetric orbifold as well: T-duality in the $X^i$ direction takes $X^i_R \rightarrow -X^i_R$, and T-dualizing both legs of a $T^2$ takes $\tau \rightarrow -1/\tau$ and $\rho \rightarrow -1/\rho$.\footnote{This is especially easy to see for a rectangular torus with sides of length $L_1$ and $L_2$, and no $B$-field. Writing $\tau = iL_1/L_2$ and $\rho = i L_1 L_2$, one can see that taking $L_{1,2} \rightarrow 1/L_{1,2}$ has the proposed action on $\tau$ and $\rho$.} Since we can undo the $\tau$ twist with a rotation of the torus, it is clear that $\rho \rightarrow -1/\rho$ is simply two T-dualities and a $90^\circ$ rotation. As such, it is an asymmetric orbifold. And, since it is T-dual to a consistent orbifold, it must also be consistent. 

This is a modular invariant asymmetric orbifold, which is certainly a nice thing to have. However, this is a fairly boring one, since it is dual to a symmetric orbifold. As we do not generally expect a nongeometric background to be dual to a geometric one, it would be nice to have a ``truly" asymmetric orbifold. The obvious next example to try, $\tau \rightarrow -1/\tau$ and $\rho \rightarrow -1/\rho$ (which is just two T-dualities) is actually {\bf not} modular invariant! There are sometimes ways of patching up such orbifolds, however, as with the more general procedure considered in \cite{Hellerman:2006tx}. For our purposes here, we will merely note that the modular consistency of asymmetric orbifolds is delicate, and that one must be careful when working with these theories. 

Another method of describing nongeometric backgrounds on the worldsheet is due to Hull \cite{x2}. Consider a torus $T^n$ fibered over some base which includes a noncontractible loop. Since the perturbative symmetry group of a $T^n$ compactification is $O(n,n;\bZ)$, we can consider fibering the torus in such a way that the transition functions between different patches are general elements of $O(n,n;\bZ)$; such compactifications are called ``T-folds" or (sadly) ``monodrofolds."  Since this group contains T-duality transformations (as well as B-field shifts and basis changes), such fibrations will generically be nongeometric. 

Hull's idea is to describe the worldsheet theory of such a fibration by doubling the number of degrees of freedom along the fiber, and then projecting back to a critical string theory by means of a constraint. More precisely, Hull promotes the fiber to a ``doubled torus" $T^{2n}$, where the extra $n$ coordinates are essentially the T-dual of the original $T^n$. We can now just consider a geometric space where this $T^{2n}$ is fibered over the base with general transition functions in $O(n,n;\bZ)$, which acts linearly on the coordinates of the $T^{2n}$. To get back to a critical string theory, we locally project onto a $T^n$ subspace of the $T^{2n}$ via a self-duality constraint. Generically this background will be nongeometric, since the transition functions between $T^n$ subspaces on different patches will involve T-duality transformations. 

One of the main utilities of Hull's approach is that one can define a worldsheet theory in the standard way, writing the worldsheet Lagrangian
\beq
	\cL = - \half \cH_{IJ}(Y) \eta^{\alpha\beta}\partial_\alpha\bX^I
	\partial_\beta \bX^J  
		- \eta^{\alpha\beta} \cJ_{IA}(Y)\partial_{\alpha}\bX^I\partial_{\beta}Y^A 
		+ \cL_N(Y),
\eeq
where $\eta_{\alpha \beta}$ is a flat worldsheet metric, $ {\mathbb X}^I$ is a coordinate on the doubled torus, $Y^A$ is a coordinate on the base, ${\cal H}_{IJ}(Y)$ is a metric on $T^{2n}$ that can depend on the base $Y$, and $\cJ_{IA}$ is a metric with mixed base-fiber terms. $\cL_N(Y)$ is the part of the Lagrangian that lives entirely on the base. In order for this theory to be equivalent to a standard sigma model, we must impose some restrictions, e.g. that the metric $\cH_{IJ}$ is a coset metric for $O(n,n)/O(n) \times O(n)$. In a particular basis, we can write the metric $\cH_{IJ}$ in terms of a given metric and B-field:
\beq
\cH = 
\left(
\begin{array}{cc}
G - B G^{-1}B  & BG^{-1}     \\
 -G^{-1}B & G^{-1}   \\
\end{array}
\right),
\eeq
after splitting the coordinates of the $T^{2n}$ as $\bX^I \rightarrow (X^i, \widetilde X_i)$. $X^i$ are the coordinates on a $T^n$ subspace, and $\widetilde X_i$ are the coordinates on the dual $T^n$. 

We will not delve further into the doubled torus formalism here, except to note that it has been a topic of recent research interest \cite{othertfold}, and it has been extended to the superstring \cite{supertfold} and demonstrated to be quantum mechanically equivalent to the usual worldsheet formulation of string theory \cite{supertfold,quantumtfold}. Additionally, it has been used to perform the mysterious third T-duality which produces the $R^{abc}$ fluxes \cite{Dabholkar:2005ve}. 

\section{Conclusions}

By this point, hopefully the reader is convinced that nongeometric compactifications are an important (and unavoidable) part of string theory. In particular, any discussion of the landscape of string vacua is incomplete without including them. Indeed, one naively expects nongeometric theories to make up the overwhelming majority of string compactifications, and any serious discussion of vacuum statistics must take them into account. 

Possible directions for future work are scattered throughout these notes, but let us here discuss some things we have not yet had a chance to mention. Since the study of nongeometric backgrounds is still in many ways in its infancy, there are many possible projects. Perhaps the most interesting open problem is to more fully grok the nature of the $R^{abc}$ flux, which is still very poorly understood. There is an increasing convergence of evidence that such backgrounds must exist, however, and it would be nice to have a better understanding of what kind of space these structures describe. 

Another pressing problem is to extend this work to non-toroidal manifolds, as begun in \cite{cts}. It is important to understand how one can incorporate nongeometric structures on Calabi-Yau manifolds, since we expect such spaces generically to appear via mirror symmetry. A big step in this direction, which we have avoided almost entirely in these notes, has been made by the study of $SU(3)$-structure and $SU(3) \times SU(3)$-structure compactifications and their duals. It appears that a generic $SU(3) \times SU(3)$-structure compactification is nongeometric, as discussed in \cite{Grana:2006hr}. It is generally unclear, however, what the overlap is between $SU(3) \times SU(3)$-structure compactifications and the models discussed in these notes, and this is something that needs to be settled. 

Finally, we mention that a general worldsheet description of the nongeometric backgrounds presented in these notes is still lacking, even just for the NS-NS sector. Although there have been several steps in this direction, it remains unclear (at least to this author) that the worldsheet techniques discussed in Section 7 are enough to account for the NS-NS backgrounds we discuss here. It is absolutely essential that we be able to understand these backgrounds via the worldsheet, if the study of nongeometric backgrounds is to be made a rigorous part of string theory.

\begin{center}
\bf{Acknowledgements}
\end{center}
\medskip
I would like to thank the organizers of the 2007 RTN Winter School for the opportunity to give these lectures, and the attendees of that school for listening. Additionally, I would especially like to thank Wati Taylor and Jessie Shelton, my collaborators on the papers on which these lectures are mostly based. This work was supported by NSF grant PHY-00-96515.

\center{These lectures are dedicated to the memory of Bonnie M. Wecht, 1942-2006.}

\end{document}